\documentclass{aa}  

\usepackage{graphicx}
\usepackage{txfonts}
\usepackage[colorlinks]{hyperref}
\usepackage{xspace}
\usepackage{mathtools}
\usepackage{amsmath}
\usepackage{gensymb}
\usepackage{lineno}

\def\flat{\textit{Fermi}/LAT\xspace}
\def\hess{H.E.S.S.\xspace}
\def\xmm{\textit{XMM-Newton}\xspace}
\def\j1702{HESS~J1702-420\xspace}
\def\ja1702{HESS~J1702-420A\xspace}
\def\jb1702{HESS~J1702-420B\xspace}
\def\ero{\textit{eROSITA}\xspace}

\newcommand{\referee}[1]{#1}
\begin{document}

   \title{Search of extended emission from \j1702 with eROSITA}

   \author{Denys Malyshev\inst{1}
          \and
          Maria Chernyakova\inst{2,3}
          \and
          Felix Aharonian\inst{3,4,5}
          \and
          Andrea Santangelo\inst{1}
          }

   \institute{Institut f{\"u}r Astronomie und Astrophysik T{\"u}bingen, Universit{\"a}t T{\"u}bingen, Sand 1, D-72076 T{\"u}bingen, Germany
         \and
            School of Physical Sciences and Centre for Astrophysics \& Relativity, Dublin City University, Glasnevin, D09 W6Y4, Ireland 
        \and
        Dublin Institute for Advanced Studies, School of Cosmic Physics, 31 Fitzwilliam Place, Dublin 2, Ireland
        \and
        Max-Planck-Institut f\"ur Kernphysik, Saupfercheckweg 1, 69117 Heidelberg, Germany
        \and
        Yerevan State University,  1 Alek Manukyan St, Yerevan 0025, Armenia }


   \date{\today}

 
  \abstract
   {\j1702 is a peculiar TeV complex with a morphology changing from a diffuse (\jb1702 source) at $\lesssim 2$~TeV to point-like (\ja1702) at $\gtrsim 10$~TeV energies.
   The morphology and the spectral properties of \j1702 could be understood in terms of a (diffusive) hadronic or leptonic models in which the observed TeV emission arises correpondingly from proton-proton or IC-radiation of relativistic particles present in the region.
   }
   {In this work we perform searches of the X-ray counterpart of \jb1702 source originated from the synchrotron emission of the primary or secondary relativistic electrons produced within leptonic or hadronic models. Such an emission can be \referee{extended} and remain beyond the detection capabilities of a narrow-FoV instruments such as \xmm.
   }
   {We utilise the publicly available first 6-months \ero dataset (DR1) fully covering selected for the analysis region of $>5^\circ$-radius around \j1702. We discuss biases connected to variable plasma temperature/neutral hydrogen column density in the region and present results based on background modelling approach. }
   {The performed analysis does not allow us to detect the extended X-ray counterpart of \j1702 of $0.07^\circ-3^\circ$-radii sizes. The derived upper limits are significantly higher than the expected hadronic model flux of the X-ray counterpart. For the leptonic model the derived limits indicate the magnetic field in the region $B\lesssim 2\,\mu$G. We argue, that the further advances in the diffuse X-ray counterpart searches could be achieved either with next generation missions or Msec-long observational campaigns with currently operating instruments.}
   {}

   \keywords{gamma-ray sources: \j1702 --
                diffusion --
                \ero
               }

   \maketitle
%

\section{Introduction}
\j1702 is a very-high energy (VHE) complex detected by High Energy Spectroscopic System (\hess) in the TeV energy band during the first Galactic plane survey campaign~\citep{hgps1}. The complex was further studied in~\citet{aharonian08}, which reported the extended morphology of the source and the first measurements of its spectral parameters. 

The recent \hess observations~\citep{hess_j1702} suggest that the morphology of the complex is changing with energy. While at low energies ($\lesssim 2$~TeV) the source is clearly characterised by a diffuse morphology, at highest energies ($\gtrsim 10$~TeV) the morphology of the source is consistent with a point-like. Such energy-dependent morphology is consistent with a superposition of a point-like source \ja1702 with a hard $\Gamma_A=1.53\pm0.2$ power law spectrum  and a much softer $\Gamma_B=2.62\pm0.2$  extended ($\sim 0.3^\circ$) source \jb1702  with an elliptical morphology. 

Peculiarly, no direct counterparts of these sources were detected at lower energies. The dedicated analysis of 10-900~GeV \flat data reported in~\cite{hess_j1702} resulted only in upper limits on the fluxes of both, extended and point-like components. \xmm observations of \j1702 region did not result in a detection of X-ray counterpart of \ja1702 source, but allowed~\citet{xmm22} to put constraints on its flux. We note that the size of \jb1702 exceeds the field of view (FoV) of \xmm which prevented \citet{xmm22} from constraining this source.

While, as suggested by~\citet{hess_j1702}, \ja1702 and \jb1702 can have independent origin, in our recent paper~\citep{we_j1702} we suggested that both sources are manifestations of a single phenomena. To explain the TeV emission from \j1702 complex we proposed a model in which VHE protons are accelerated in (or close to) point-like source \ja1702 and injected into a dense surrounding neutral hydrogen's cloud of characteristic size of $\sim 0.3^\circ$ and density $n_0\sim 100$~cm$^{-3}$. The observed emission originate from pion-decay process and observed morphological changes correspond to the transition in protons' propagation regime. The low-energy protons are propagating in a strongly diffusive regime, and consequently characterised by quasi-isotropic angular momenta distribution. The pion decay emission from protons propagating in this regime result in an extended TeV source \jb1702. To the contrary, the high-energy protons are propagating in a (quasi) ballistic regime and characterised by delta-function like distribution of angular momenta. In this regime the observed $\gamma$-ray emission is characterised by a point-like morphology. A similar model was proposed earlier by~\citet{Chernyakova2011} to describe the spectral behaviour of the Galactic Center in the GeV energy band.
\begin{figure*}
    \centering
    \includegraphics[width=0.95\textwidth]{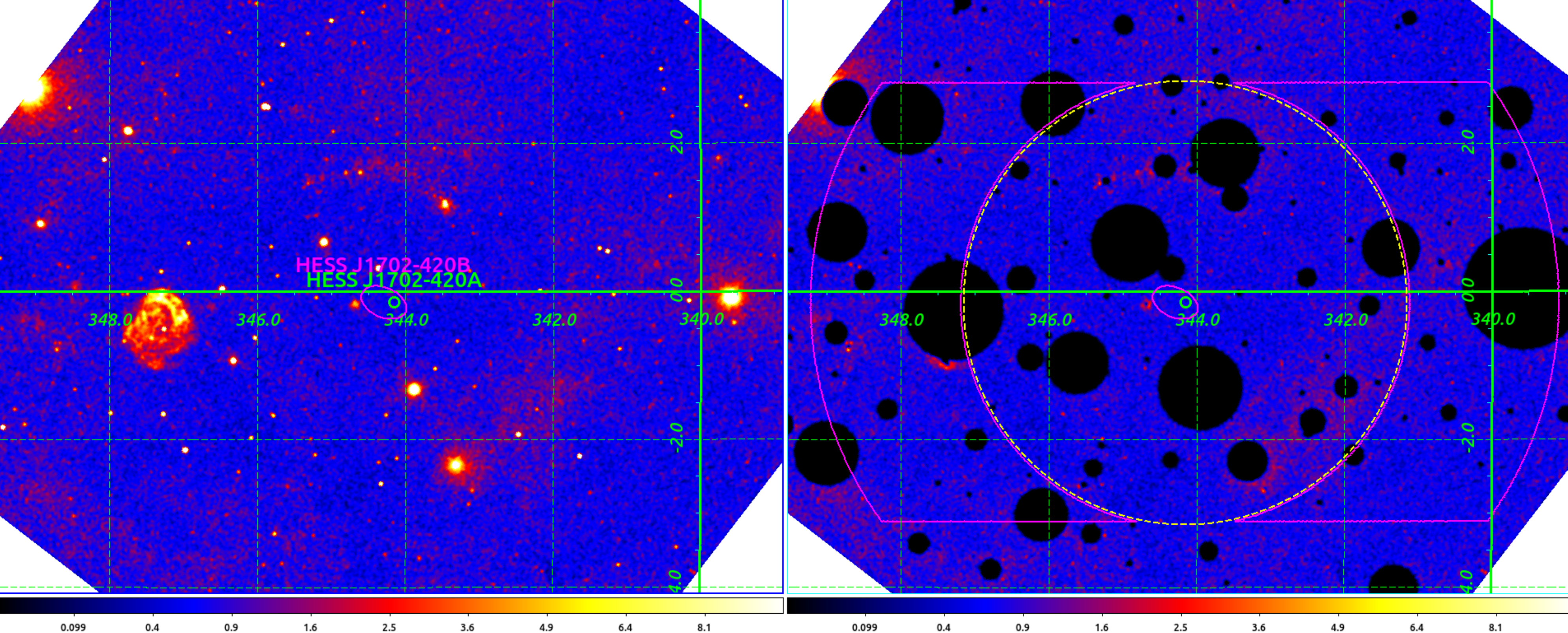}
    \caption{\ero count map of \j1702 region (in galactic coordinates). Green and magenta ellipses in both panels indicate positions and TeV sizes of \ja1702 and \jb1702 sources. Right panel shows the region with the bright nearby X-ray sources masked for the searches of the diffuse emission from X-ray counterpart of \j1702 TeV complex. The dashed yellow circle of $3^\circ$-radius is centered at \ja1702 position and illustrates the maximal radius of ON region used for the searches. Magenta cutted annulus in the right panel illustrates the shape and the extent of the corresponding OFF region.}
    \label{fig:sky_map}
\end{figure*}

To explain simultaneously the observed spatial and spectral behaviour of \ja1702 and \jb1702 sources \citet{we_j1702} proposed that the region is characterised by a diffusion coefficient $D(E_p) = D_0(E_p/1\mbox{TeV})^\beta$, which is essentially suppressed at low energies in comparison to the interstellar medium ($D_0 \sim 10^{26}$~cm$^2$/s at 1 TeV) but with a strong energy-dependence ($\beta \gtrsim 1$) that results in the propagation of highest energy protons ($E \geq 100$~TeV) in ballistic regime. The detected fluxes of gamma rays require a powerful proton accelerator with an injection rate at the level of  $Q_0 \sim 10^{38}(n_0/100 \mbox{cm}^{-3})^{-1}$~erg/s.

The secondary electrons produced as a result of the pion-decay process simultaneously with TeV photons are expected to emit the synchrotron radiation, potentially detectable in the X-ray band. This radiation is expected to be characterised by a hard ($\Gamma \sim 2$) X-ray spectrum and a flux $F_0\sim (2..3)\cdot 10^{-13}$~erg/cm$^2$/s at 1~keV for a broad range ($10-100\,\mu$G) of considered magnetic fields in the region~\citep{we_j1702}. The spatial scale of the emission is expected to significantly exceed the size of \j1702 complex and be of a degree-scale for preferable ($\sim$kpc) distances to this complex.

Alternatively, the sources can have an independent origin with the TeV emission produced by a population of relativistic electrons present in the region and emitting Inverse Compton (IC) on a soft-photons radiation field~\citep{hess_j1702, xmm22}. A search of the X-ray counterpart for a point-like \ja1702 source for this-type model was performed by~\citet{xmm22} with \xmm. This study did not result in a significant detection of point-like X-ray counterpart, but allowed to constrain magnetic field in \ja1702 vicinity to be $\lesssim 1.5\,\mu$G. At the same time the narrow field of view (FoV) of \xmm did not allow \citet{xmm22} to perform searches for the counterpart of a diffuse source \jb1702.

In this work we perform a search for an extended emission around \ja1702 position, as expected X-ray counterpart of \j1702 complex. For these purposes we utilise data-release DR1 dataset~\citep{dr1} of \ero satellite~\citep{erosita}, i.e. the data taken within first 6 months of \ero operation. The publicly available data fully covers the western part of the Galactic hemisphere~\citep{dr1} and thus is perfectly suitable for searches of the large-scale diffuse emission.

\begin{figure*}
 \centering
    \includegraphics[width=0.48\textwidth]{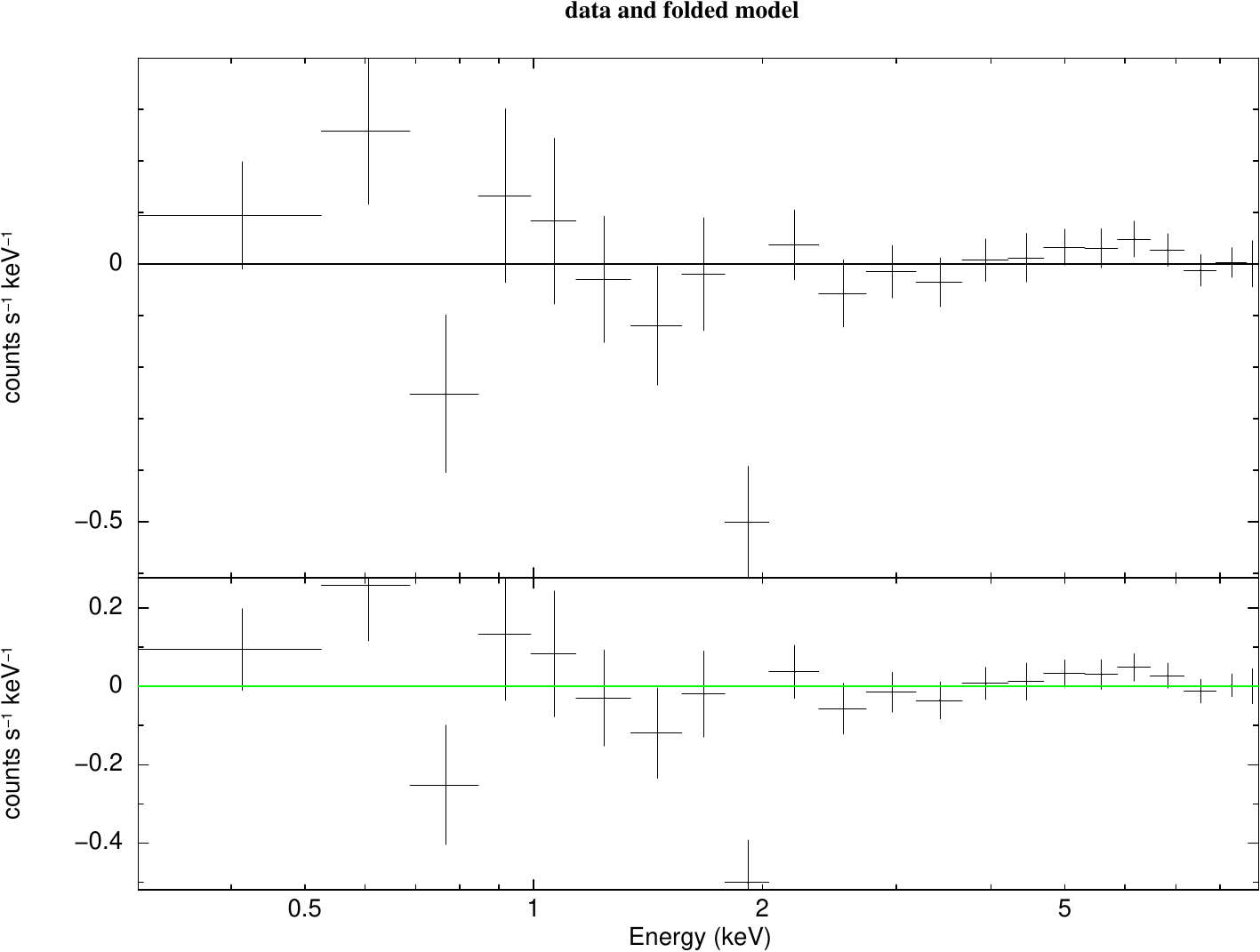}
    \includegraphics[width=0.48\textwidth]{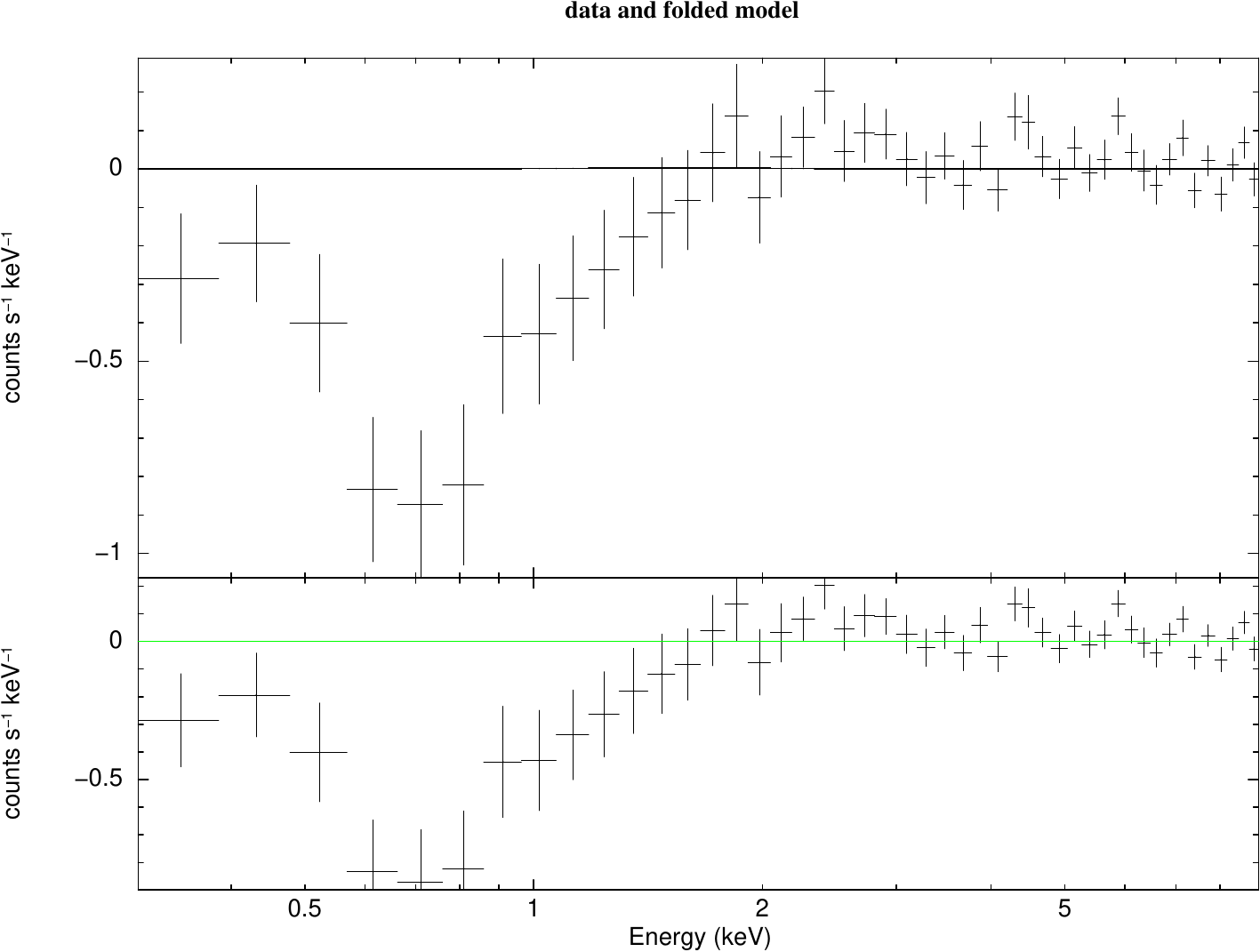}
    \caption{ ON-OFF (background subtracted) \ero spectra of $0.5^\circ$-radius (left panel) and $1^\circ$-radius (right panel) regions centered at \ja1702 with the residuals with respect to the best-fit absorbed powerlaw model (bottom sub-panels). Right panel demonstrate the bias of the subtraction procedure due to the changes of the $n_H$ between ON and OFF regions, see text for the details and Fig.~\ref{fig:spectrum_model}.}
    \label{fig:spectra_on_off}   
\end{figure*}

\section{Data Analysis}
\label{sec:analysis}
For the analysis we utilised all publicly available \ero DR1 data (data processing version 010), analysed it with \texttt{eSSAS4DR1} and \texttt{heasoft}-6.31 and performed spectral analysis with XSPEC v.12.13.0c software. We selected for the analysis 19 \ero sky tiles\footnote{Available for download from \href{https://erosita.mpe.mpg.de/dr1/erodat/skyview/skytile_search/}{\ero website} } overlapping with $5^\circ$-radius circle around \ja1702 position. 

At the initial step of our analysis\footnote{Please see \href{https://erosita.mpe.mpg.de/dr1/eSASS4DR1/eSASS4DR1_cookbook/}{\ero analysis cookbook} for the details} we build the count map of the analysed region in 0.3-10~keV energy band with the help of \texttt{evtool} routine. Corresponding map is shown in the left panel of Fig.~\ref{fig:sky_map} overlaid with green and magenta ellipses illustrating positions and spatial extent in the TeV band of \ja1702 and \jb1702 sources correspondingly. Aiming in searches of the potentially largely spatially-extended (suggested to be of a degree-scale~\cite{we_j1702}) emission from the X-ray counterpart(s) of these sources we explicitly masked for the analysis all bright X-ray sources detected in the region. Namely, we performed a search of sources detected at up to $\gtrsim 3\sigma$ level above the local background with the help of \texttt{erbox} and created a mask of the region, masking all sources above detection likelihood threshold equal to 5 with the help of \texttt{erbackmap} routines. The count map of the analysed region with the applied exclusion mask is shown in Fig.~\ref{fig:sky_map}, right panel.

We note, that there are no visible indications of presence of a large spatial-scale residual emission\footnote{Note, that the diffuse shell-like structure at $(\ell,b)=(347.3^\circ, -0.5^\circ)$ is a known GeV/TeV SNR HESS~J1713.7-3946~\citep{hessj1713}. } centered in the vicinity of \j1702 in the Fig.~\ref{fig:sky_map}. In order to justify this qualitative statement we performed a dedicated spectral analysis aiming either to detect the extended emission in the region or to put constraints on its flux.

\subsection{Hadronic model: secondary electrons' synchrotron emission}
\label{sec:data_analysis_hadronic}
In this subsection we explicitly assume that the X-ray emission originates from the synchrotron emission of the secondary electrons produced in hadronic model~\citet{we_j1702}. In this case the expected X-ray signal is characterised by a hard spectrum $dN/dE\propto E^{-2}$~\citep{we_j1702} and the described below analysis was performed for this spectral index.

For the spectral analysis we selected an ``ON'' region of a certain radius $R_{ON}$ (see Tab.~\ref{tab:spectral_params}) centered at the position of \ja1702 ($\ell=344.15^\circ$, $b=-0.15^\circ$) and corresponding to the possible size of the X-ray counterpart of the TeV source \j1702. For the estimations of the background flux we utilised the ``OFF''-region of the ``cutted annulus'' shape. The borders of the ``OFF''-region are given by an annulus centered at \ja1702 position with the inner radius $R_{in}=R_{ON}$ and outer $R_{OFF}=1.7R_{ON}$. Aiming to minimize the impact of possible background variations due to somewhat different average galactic latitudes of ON and OFF regions we additionally cut the OFF-region annulus at minimal/maximal galactic latitudes of the ON region. The characteristic shapes of ON and OFF regions are illustrated in the right panel of Fig.~\ref{fig:sky_map} with yellow dashed circle and magenta cutted annulus correspondingly. The exposures of the spectra varies from $\sim 0.3$~ksec for $R_{ON}=0.07^\circ$ to $\sim 10$~ksec for $R_{ON}=3^\circ$ region.

The widely used ``ON-OFF'' approach for the estimation of the presence of the signal in ON region relies on the assignment of the OFF-region spectrum as a background for the ON-spectrum and subsequent fitting of the resulted residual spectrum with a certain model. However such an approach relies strongly on the assumption of absence of the variations of the parameters of the background emission in ON and OFF regions. Such an assumption, however, could fail in case if ON and/or OFF regions are significantly extended. In this case the astrophysical background emission in ON and OFF region could be characterised by somewhat different parameters of the medium (e.g. neutral hydrogen column density and/or temperature of the interstellar plasma) which results in a biased (over or under-subtracted) residual ON-OFF spectrum.

We demonstrate the presence of such an issue with the ON-OFF analysis of the \j1702 vicinity for the regions of characteristic scale of $\gtrsim 1^\circ$-radius. Fig.~\ref{fig:spectra_on_off}, left panel illustrates the ON-OFF (background subtracted) spectrum for the $R_{ON}=0.5^\circ$-radius size of the ON region (top sub-panel) and residuals with respect to the best-fit absorbed powerlaw model of the signal ( \texttt{tbabs*powerlaw} in terms of XSPEC\footnote{With ``wilms'' abundances~\citep{wilms_abund}} with index fixed to 2). The residual spectrum is statistically consistent with 0, does not demonstrate any energy-depended features and can be used to derive the unbiased upper limits on the normalisation of the signal's model. The right panel of Fig.~\ref{fig:spectra_on_off} shows similar ON-OFF spectrum for the $R_{ON}=1^\circ$. In this case the spectrum is characterised by a wiggle-like structure with strong over-subtraction at $\lesssim 2$~keV energies and under-subtraction in $\sim 2-3$~keV energy range. The shown ON-OFF spectrum is unsuitable thus for an unbiased estimations of the strength of the signal. We argue, that the observed wiggle-like structure is connected to variations of the neutral hydrogen density and/or interstellar plasma temperature between ON and OFF regions, i.e. on $\sim 1^\circ$ spatial scales.

Thus, to derive the robust constraints on the flux of the possible spatially-extended source present in the region instead of ON-OFF approach we adopted a background-modelling technique, widely used e.g. in the dark matter searches~\citep[see e.g.][]{we_ths, we_pbh}. Namely, instead of considering a difference of ON and OFF spectra we focus on a joint spectral modelling of these quantities. We propose to model spectra of ON and OFF regions with a sum of the astrophysical and instrumental background models. 

To the model of astrophysical background we included the
following components: \textit{(i)} powerlaw with the spectral index fixed to 2 (presenting the possible signal from X-ray counterpart of \j1702); \textit{(ii)} low-temperature plasma ($kT\sim 0.2$~keV, presenting the contribution from the plasma local to the Solar system); \textit{(iii)} high-temperature plasma ($kT\sim 0.8$~keV, presenting the hot plasma in the MW); \textit{(iv)} powerlaw with the index fixed to 1.4 (presenting the contribution from cosmic X-ray background~\citep{luca04}). We additionally corrected the described model for the interstellar neutral hydrogen absorption. In terms of XSPEC the model is given by 
\texttt{ apec + tbabs(powerlaw + apec + powerlaw)  } model. Please note, that the \texttt{apec} model component not convolved with the interstellar absorption present the low-temperature plasma local to the Solar system. 

To the model of the instrumental background we include a broken powerlaw and 3 narrow gaussian lines at energies $\sim 0.9$~keV (Ne IX line); $\sim 1.8$~keV (Si K$\alpha$/Si K$\beta$ lines); $\sim 6.4$~keV (Fe K$\alpha$ line), see e.g.~\cite{we_pbh} for a summary of the typical \xmm instrumental lines. Note, that the instrumental model was not convolved with \ero effective area.

\begin{figure}
    \centering
    \includegraphics[width=0.95\columnwidth]{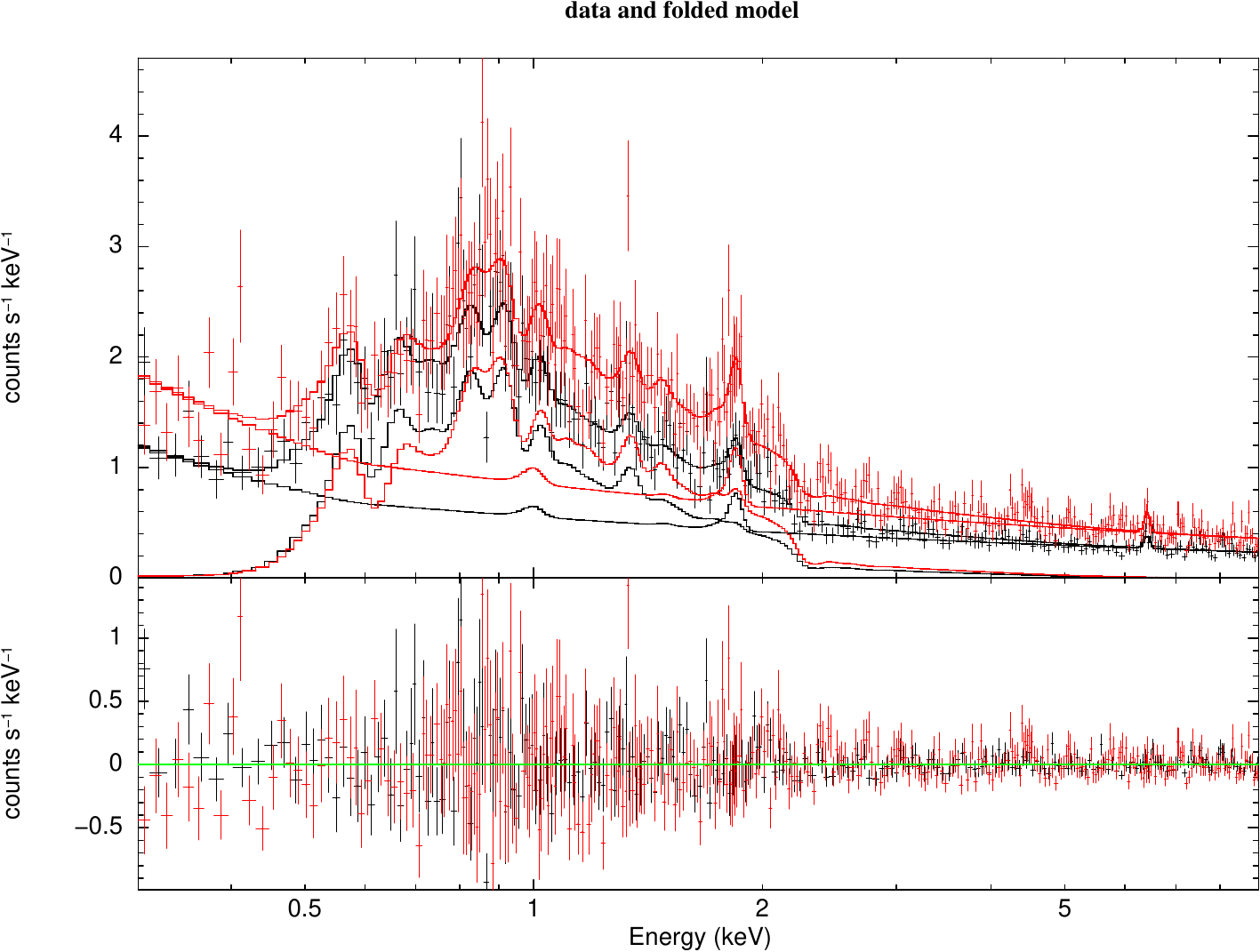}
    \caption{\ero spectrum of $1^\circ$ region. Red points and curve illustrate the spectrum and best-fit model for the ON region, black points and curves -- the same for OFF region. Bottom panel illustrate residuals with respect to the selected models. }
    \label{fig:spectrum_model}
\end{figure}
To derive the constraints on the flux of the extended source we simultaneously modeled the spectra of ON and OFF regions with a sum of described models. We additionally required that \textit{(i)} the temperature of the local plasma is identical for ON and OFF regions; \textit{(ii)} the flux of the astrophysical components of the model in ON and OFF regions is proportional to the not-masked area of the analysed region (basing on BACKSCALE keyword of the corresponding spectra). The flux from the possible X-ray counterpart of \j1702 was explicitly set to 0 in OFF region. The characteristic spectra of ON (red points and curves) and OFF (black points and curves) regions along with the proposed best-fit models and residuals with respect to these models are shown in Fig.~\ref{fig:spectrum_model}.

The best-fit values of the key parameters of astrophysical model for ON and OFF regions are summarized in Tab.~\ref{tab:spectral_params}. This Table summarizes the temperature of the plasma local to the Solar system ($kT_{sol}$); temperatures of the interstellar plasma in ON and OFF regions ($kT_{ON}$ and $kT_{OFF}$ correspondingly); neutral hydrogen column density for ON and OFF regions ($n_{H,ON}$ and $n_{H,OFF}$ correspondingly) and the limit on the flux of an extended source with $dN/dE\propto E^{-2}$ spectrum at 1~keV energy. The flux limit correspond to $2\sigma$ ($\sim 95$\% c.l.) statistical limit and was calculated with \texttt{error 4.0} XSPEC command.
The columns $R_{ON}$ and $R_{OFF}$ indicate the radius of the ON region and the outer radius of the OFF region correspondingly.

For the relatively small radii of the ON region the statistics of the data does not allow firm measurement of the plasma temperatures and/or interstellar absorption. In these cases we fixed corresponding parameters to the values indicated in the Tab.~\ref{tab:spectral_params} basing on the values observed at larger $R_{ON}$. For the smallest $R_{ON}=0.07^\circ$ corresponding to the size of H.E.S.S. point spread function the statistics of the data does not allow any firm measurement of the neutral hydrogen absorption. In this case we explicitly consider two $n_H$ values -- $0.7\cdot 10^{22}$\,cm$^{-2}$ as indicated by best-fit $n_H$ values at larger $R_{ON}$ and $1.7\cdot 10^{22}$\,cm$^{-2}$ as suggested by mean galactic $n_H$ value in \j1702 direction\footnote{As reported at \href{https://heasarc.gsfc.nasa.gov/cgi-bin/Tools/w3nh/w3nh.pl}{heasarc nH column density website}}. For this $R_{ON}$ value we indicate in the Table~\ref{tab:spectral_params} two values for the flux limit and use in what follows the conservative limit.

The described models fit the data with the reduced $\chi^2\lesssim 1$ for all $R_{ON}<2^\circ$. For lager radii we noticed the worsening of the fit possibly originating from the presence of multi-temperature and/or variable $n_H$ sub-regions within analysed area. To derive robust constraints in these cases we consider \referee{``systematics-driven approach''. We increased the systematics (added in quadratures), up to a level required to make the reduced $\chi^2$ of the fit to be 1. The characteristic values of systematics in this case were $\sim 5$\%, see ``Systematic'' column of Tab.~\ref{tab:spectral_params}.}

\begin{figure}
    \centering
    \includegraphics[width=0.95\columnwidth]{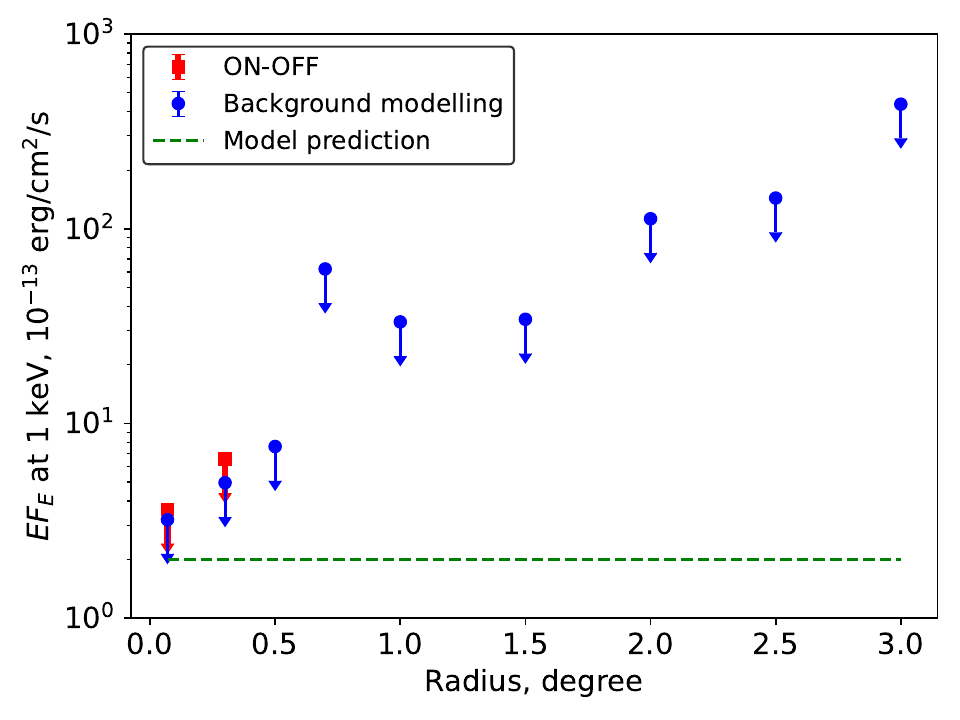}
    \caption{\ero limits on \j1702 flux at 1 keV based on \ero-DR1 data. Green line illustrate the model-predicted flux from \citet{we_j1702}. 
    }
    \label{fig:flux_limits}
\end{figure}

The derived upper limits on the flux at 1~keV of the extended source centered at the position of \ja1702 as a function of the radius of the source are shown in Fig.~\ref{fig:flux_limits}. The red square points indicate results from ON-OFF analysis performed at $R<0.5^\circ$ where the bias caused by different plasma temperatures/$n_H$ values in ON and OFF regions can be neglected. Blue circle points present the results from background modelling approach and correspond to flux limit values indicated in Tab.~\ref{tab:spectral_params}. The green dashed line illustrate the characteristic flux of the extended source expected from \citet{we_j1702} model. We note also a weak, statistically insignificant detection at $\sim 1.5\sigma$ level of the signal at $R_{ON}\sim 0.7^\circ$ which somewhat weakened the flux limits at these radii.

\referee{We note that} the dedicated analysis limited only to cameras not affected by a light-leak\footnote{See e.g. \href{https://erosita.mpe.mpg.de/edr/eROSITAIssues/lightleak.html}{\ero known issues webpage}} results in very similar upper limits on the flux of the extended source.

\subsection{Leptonic model: primary electrons' synchrotron emission}
\label{sec:data_analysis_leptonic}
As an alternative to the hadronic model described in \citet{we_j1702} we consider a simple leptonic model. In this model the observed diffuse TeV emission from \jb1702 is explained as an Inverse Compton emission of a powerlaw population of electrons producing VHE emission on the CMB photons. The emission in the X-ray band in this model corresponds to the synchrotron emission from the same population electrons. The emission of the discussed model components was calculated with \texttt{naima} v.~0.10.0 model~\citep{naima} which uses \citet{aharonian10, khangulian14} approach to calculate synchrotron and IC components correspondingly. 

We derive the parameters of the relativistic electron population by fitting the described IC radiation model to the \hess data on \jb1702. We found that this population is characterised by a relatively soft spectrum of $\Gamma_e=3.85\pm 0.15$ and the total energy in electrons above 1~TeV is $W_e=2.6\cdot 10^{46}$~erg, see Fig.~\ref{fig:leptonic model}. The corresponding X-ray spectrum is characterised by an index $\Gamma\sim 2.4$ and normalisation proportional to the square of strength of the magnetic field in the region.

Assuming that the X-ray signal originates from a region identical to the spatial shape of \jb1702 source\footnote{An ellipse, centred at $(\ell, b)=(344.29^\circ, -0.15^\circ)$, semi-axes $(0.32^\circ, 0.2^\circ)$ and rotation angle $157^\circ$} we have performed an analysis similar to described in Sec.~\ref{sec:data_analysis_hadronic}. Given relatively large size of the signal-extraction region we focused on the background modelling approach. Namely, we extracted background spectrum from the elliptic-annulus region with the inner boundaries coinciding with the signal extraction region. The outer boundary of the background region was defined as an \referee{ellipse a factor of 1.5 larger than the signal region.}

We found, that similar to Sec.~\ref{sec:data_analysis_hadronic} the data is well-fitted by the described model with zero-level systematic uncertainty. The 95\% c.l. upper limit on the flux at 1~keV for the discussed spectral index is $\sim 2.5\cdot 10^{-4}$~ph/cm$^2$/s or $\sim 4\cdot 10^{-13}$~erg/cm$^2$/s (red upper limit in Fig.~\ref{fig:leptonic model}). \referee{This upper limit on the flux} corresponds to the \referee{95\%} upper limit on the magnetic field in the considered region of $B\lesssim 1.5\,\mu$G.  

In order to estimate the potential impact of systematic uncertainties on this value we have considered an alternative model, in which the IC TeV emission is produced on Interstellar Radiation Field background~\cite{isrf} (for $R=8$~kpc ISRF model). We found that for this model the upper limit on the magnetic field is somewhat relaxed to $B\lesssim 1.7\,\mu$G. We note, that for the models suggesting higher density of the soft photons' background one can expect even more relaxed limits on the magnetic field.

An additional source of systematic uncertainty could arise from the systematic uncertainty on the flux level of \jb1702 measured by \hess that could reach $\sim 20$\%~\citep{hess_crab}. This uncertainty correspond to an additional $\sim 10$\% systematic uncertainty on the derived limit on the magnetic field in the region, putting it at a level of $\lesssim 2\,\mu$G.


\begin{table*}
    \centering
    \begin{tabular}{|c|c||c||c|c||c|c||c|c|c|}
    \hline
     $R_{ON}$ & $R_{OFF}$ & $kT_{sol}$  &  $kT_{ON}$ & $kT_{OFF}$ & $n_{H,ON}$ & $n_{H,OFF}$ & Flux limit& Syst. & $\chi^2/ndf$\\ 
     \small degree &\small degree & keV  &  keV & keV & $10^{22}$\,cm$^{-2}$ & $10^{22}$\,cm$^{-2}$ &\small ph/keV/cm$^2$/s & &\\ \hline
     0.07 & 0.12 & 0.17*  &  0.68* & 0.68* & $0.7-1.7$* & $=n_{H,ON}$  & \small$(0.8-2)\cdot 10^{-4}$ & 0 &10.8/16\\
     0.3  & 0.51 & $0.17\pm 0.025$& $0.68\pm 0.08$ & $=kT_{ON}$ & $0.7\pm 0.4$ & $=n_{H,ON}$  & $3.1\cdot 10^{-4}$& 0 & 292/286\\ 
     0.5 & 0.85 & $0.18\pm 0.03$ &  $0.61\pm 0.06$ & $=kT_{ON}$ &  $0.77\pm 0.25$ & $=n_{H,ON}$ & $4.8\cdot 10^{-4}$ & 0 & 604/595\\
     0.7 & 1.19 & $0.17\pm 0.02$ & $0.71\pm 0.07$ & $0.66\pm 0.06$ & $0.54\pm 0.1$ & $0.33\pm 0.1$ & $3.9\cdot 10^{-3}$ & 0 & 917/917\\
     1.0 & 1.7 & $0.19\pm 0.02$&  $0.72\pm 0.05$ & $0.71\pm 0.04$ & $0.44\pm 0.06$ & $0.26\pm 0.05$ & $2.1\cdot 10^{-3}$ & 0 & 1206/1257\\
     1.5 & 2.55 & $0.22\pm 0.01$&  $0.74\pm 0.03$& $0.85\pm 0.02$& $0.53\pm 0.04$ & $0.28\pm 0.03$ & $2.1\cdot 10^{-3}$ & 0 & 1621/1564\\
     2.0 & 3.4 & $0.21\pm 0.005$&  $0.77\pm 0.02$&$0.78\pm 0.02$ & $0.39\pm 0.02$ & $0.21\pm 0.02$ & $7.0\cdot 10^{-3}$ & 0.05 & 1644/1643\\ 
     2.5 & 4.25  & $0.22\pm 0.005$ & $0.79\pm 0.02$ & $0.79\pm 0.02$ & $0.32\pm 0.02$ & $0.19\pm 0.015$ & $9.0\cdot 10^{-3}$ & 0.04 & 1636/1647\\ 
     3.0 & 5.1 & $0.22\pm 0.003$&  $0.79\pm 0.02$ & $0.82\pm 0.02$ & $0.28\pm 0.015$ & $0.23\pm 0.02$ & $2.7\cdot 10^{-2}$ & 0.04 & 1606/1650\\ \hline

    \end{tabular}
    \caption{The best-fit parameters of the astrophysical background model used to describe the data within ON and OFF regions, see text for the details. The Table summarizes the temperature of the plasma local to the Solar system ($kT_{sol}$); temperatures of the interstellar plasma in ON and OFF regions ($kT_{ON}$ and $kT_{OFF}$ correspondingly); neutral hydrogen column density for ON and OFF regions ($n_{H,ON}$ and $n_{H,OFF}$ correspondingly) and the limit on the flux at 1~keV of an extended source centered at \ja1702 position and characterised by $dN/dE\propto E^{-2}$ spectrum. The flux limit correspond to $2\sigma$ ($\sim 95$\% c.l.) statistical limit. The columns $R_{ON}$ and $R_{OFF}$ indicate the radius of the ON region and the outer radius of the OFF region correspondingly. ``Syst'' column indicate considered level of systematics \referee{and $\chi^2/ndf$ column shows the best-fit $\chi^2$ value and the corresponding number of degrees of freedom.}}
    \label{tab:spectral_params}
\end{table*}

\section{Results and Discussion}
\label{sec:discussion}

\begin{figure}
    \centering
    \includegraphics[width=0.99\columnwidth]{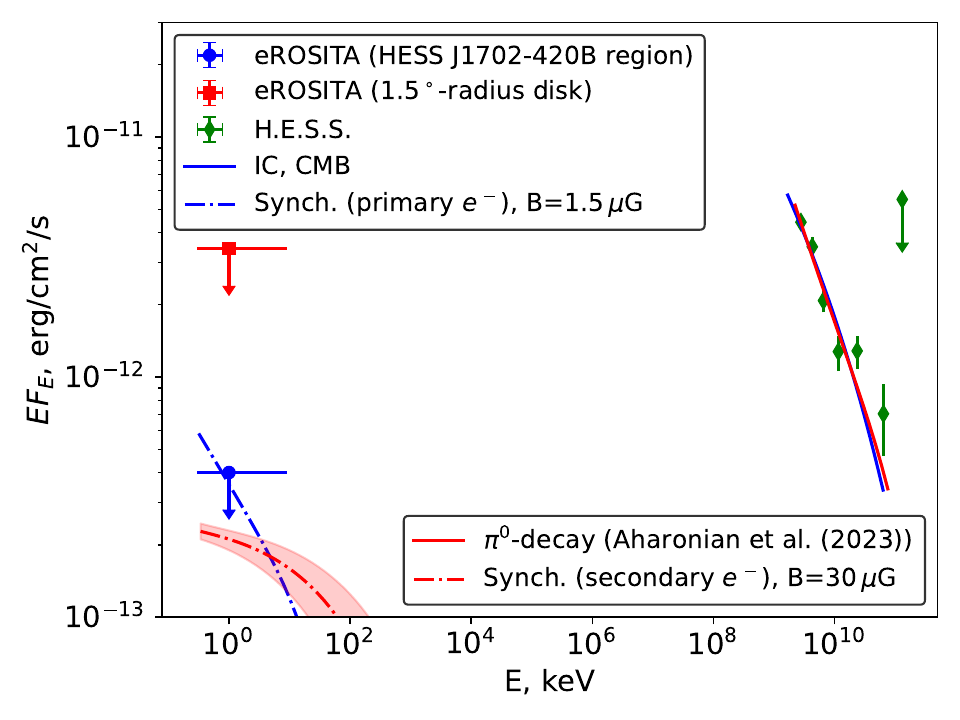}
    \caption{The \hess spectrum (green points) and \ero flux upper limit for \jb1702-shape region (blue circle upper limit) and for $1.5^\circ$-radius disk-shape (red square upper limit) region. The components of the leptonic model are shown with solid (IC) and dot-dashed (synchrotron) blue lines. The synchrotron component is shown for the 95\% c.l. excluded value of the magnetic field ($\sim 1.5\,\mu$G) see Sec.~\ref{sec:data_analysis_leptonic} for the details. The components of \citet{we_j1702} model are shown with solid ($\pi^0$-decay) and dot-dashed (synchrotron emission from secondary electrons) red lines. The shaded region illustrates the expected scatter of X-ray fluxes in \citet{we_j1702} model for $B=10-100\,\mu$G.
    }
    \label{fig:leptonic model}
\end{figure}

In this work we present the results of the search for a possible X-ray counterpart of \j1702 complex~\citep{hess_j1702}. In the TeV band it is characterised by a peculiar energy-dependent morphology and could be described as a superposition of two sources -- point-like \ja1702 and significantly extended ($\sim 0.3^\circ$) \jb1702. These sources demonstrate hard $\Gamma\sim 1.5$ (\ja1702) and soft $\Gamma\sim 2.5$ (\jb1702) powerlaw spectra up to few tens TeV without cutoff indications.

According to the model proposed by \citet{we_j1702} the morphology and spectral behaviour of \j1702 complex can be explained in terms of hadronic emission from the protons accelerated in \ja1702 source and propagating through the surrounding medium in an energy-dependent regime. The synchrotron emission from the secondary electrons produced in $\pi^0$-decay is characterised by a hard $\Gamma\sim 2$ spectrum (see Fig.~\ref{fig:leptonic model}) and could be potentially detected in the keV energy band. \citet{we_j1702} suggested that the characteristic spatial size of the X-ray counterpart of \j1702
significantly exceeds TeV-band size of \jb1702 source and reach degree-scale values.

As an alternative to~\cite{we_j1702} model we considered a simple leptonic model in which the observed diffuse TeV emission originate from an IC emission of the population of relativistic electrons on a soft radiation field present in the region. In this case the X-ray emission is expected to be characterised by a softer slope of $\Gamma\sim 2.4$ (see Fig.~\ref{fig:leptonic model}) and flux level defined by the strength of the magnetic field in the region. 

In our work we performed a search for a counterpart in \ero DR1 dataset covering first 6 months of data taking. Our analysis indicates that \j1702 is located in a relatively crowded X-ray region (see Fig.~\ref{fig:sky_map}) characterised by degree-scale gradients of the interstellar plasma temperature and/or neutral hydrogen column densities, see Tab.~\ref{tab:spectral_params}. Similarly to previous \xmm observations~\citep{xmm22} we did not find point-like X-ray counterparts of \ja1702 and \jb1702 sources within TeV error-ellipses\footnote{\referee{The closest sources are located $\sim 0.1^\circ$ and $\sim 0.25^\circ$ away from \jb1702 ellipse and can be associated with SNR G344.7-00.1 and HMXB OAO 1657-41 correspondingly.}} and focused on searches for the largely extended emission in this region. 

In order to constrain hadronic model of \citet{we_j1702} we performed a search for a hard spectrum ($\Gamma=2$), disk-like spatial morphology sources centered at the position of \ja1702. Our searches did not result in a significant detection of the emission in 0.3-10~keV energy band for source radius varying between $0.07^\circ$ and $3^\circ$. Fig.~\ref{fig:flux_limits} shows the derived flux limits at 1~keV as a function of the radius of the source. The red points illustrate the limits derived within the widely used ON-OFF, ``background subtraction'' approach, while the blue ones stand for the background modelling method. Note, that the ON-OFF approach at larger radii could result in biased limits, see Fig.~\ref{fig:spectra_on_off} and Sec.~\ref{sec:analysis} for the details. The green dashed line in Fig.~\ref{fig:flux_limits} indicate the expected flux level of the \j1702 counterpart predicted by \citet{we_j1702} model.

The derived limits are significantly weaker than the flux level expected from \citet{we_j1702} model. This does not allow us to firmly confirm or reject the proposed model with the current \ero data. While this work is based only on first six months of \ero data (DR1 data-release), about 2~years of \ero data will become public at the end of 2025\footnote{According to \href{https://erosita.mpe.mpg.de/erass/}{\ero data release schedule} }. With the same all-sky monitoring strategy this will increase the available exposure by a factor of 4, and in case of absent systematic could allow improvement of presented limits by a factor of 2. Although this improvement is not significant enough to probe the predicted by \citet{we_j1702} flux range, it could help to detect the source, if the model flux was underestimated by a factor of a few. An order of magnitude improvement of the current limits, required to reach predicted by~\cite{we_j1702} flux level from a few-degree-scale source, could be achieved (assuming statistical uncertainty only) with $\sim 100$ times longer exposure, see Fig.~\ref{fig:leptonic model}. Namely, such an improvement would require $\sim 10^6$~s long observations with \ero or similar missions. 

We note also, that our analysis indicated variations of the neutral hydrogen density along the analysed region. The density is highest close to \ja1702 position with a trend to decrease at larger distances (i.e. larger $R_{ON}$ and $R_{OFF}$ radii), see Tab.~\ref{tab:spectral_params}. This \referee{is consitent with} \cite{we_j1702} model, suggesting that the TeV emission from \jb1702 source is connected with a dense molecular cloud of $\sim 0.3^\circ$ size surrounding the VHE protons' accelerator \ja1702.

Our analysis performed for the leptonic model did not lead to the detection of the X-ray emission in the spatial region coinciding by the shape with \jb1702 source. We estimated the magnetic field strength in the region to be $B\lesssim 2\,\mu$G (including possible systematic uncertainties) which is in a broad agreement with the characteristic strength of the interstellar magnetic field~\citep{jansson_farrar}. We note, that the derived value is also consistent with the value of the magnetic field for the point-like source \ja1702 derived by~\cite{xmm22} ($B\lesssim 1.5\,\mu$G). 

\referee{We argue, that the further advances in our understanding of possible X-ray emission from \j1702 
could be reached} either with a broad-FoV, large effective area future missions e.g. Athena/WFI~\citep{athena_wfi} or Msec-long observational campaigns with the current-generation instruments (\xmm or \ero) which will allow at least an order of magnitude improvement of the presented upper limits based on ksec-long observations of \ero.

\subsection*{Acknowledgements}
The authors acknowledge support by the state of Baden-W\"urttemberg through bwHPC.

\bibliography{biblio}

\begin{thebibliography}{20}
\expandafter\ifx\csname natexlab\endcsname\relax\def\natexlab#1{#1}\fi

\bibitem[{{Abdalla} {et~al.}(2021){Abdalla}, {Aharonian}, {Ait Benkhali},
  {Ang{\"u}ner}, {Arcaro}, {Armand}, {Armstrong}, {Ashkar}, {Backes},
  {Baghmanyan}, {Barbosa Martins}, {Barnacka}, {Barnard}, {Becherini}, {Berge},
  {Bernl{\"o}hr}, {Bi}, {B{\"o}ttcher}, {Boisson}, {Bolmont}, {de Bony de
  Lavergne}, {Breuhaus}, {Brun}, {Brun}, {Bryan}, {B{\"u}chele}, {Bulik},
  {Bylund}, {Caroff}, {Carosi}, {Casanova}, {Chand}, {Chandra}, {Chen},
  {Cotter}, {Cury{\l}o}, {Damascene Mbarubucyeye}, {Davids}, {Davies}, {Deil},
  {Devin}, {Dirson}, {Djannati-Ata{\"\i}}, {Dmytriiev}, {Donath}, {Doroshenko},
  {Dreyer}, {Duffy}, {Dyks}, {Egberts}, {Eichhorn}, {Einecke}, {Emery},
  {Ernenwein}, {Feijen}, {Fegan}, {Fiasson}, {Fichet de Clairfontaine},
  {Fontaine}, {Funk}, {F{\"u}{\ss}ling}, {Gabici}, {Gallant}, {Giavitto},
  {Giunti}, {Glawion}, {Glicenstein}, {Grondin}, {Hahn}, {Haupt}, {Hermann},
  {Hinton}, {Hofmann}, {Hoischen}, {Holch}, {Holler}, {H{\"o}rbe}, {Horns},
  {Huber}, {Jamrozy}, {Jankowsky}, {Jankowsky}, {Jardin-Blicq}, {Joshi},
  {Jung-Richardt}, {Kasai}, {Kastendieck}, {Katarzy{\'n}ski}, {Katz},
  {Khangulyan}, {Kh{\'e}lifi}, {Klepser}, {Klu{\'z}niak}, {Komin}, {Konno},
  {Kosack}, {Kostunin}, {Kreter}, {Lamanna}, {Lemi{\`e}re}, {Lemoine-Goumard},
  {Lenain}, {Leuschner}, {Levy}, {Lohse}, {Lypova}, {Mackey}, {Majumdar},
  {Malyshev}, {Malyshev}, {Marandon}, {Marchegiani}, {Marcowith}, {Mares},
  {Mart{\'\i}-Devesa}, {Marx}, {Maurin}, {Meintjes}, {Meyer}, {Mitchell},
  {Moderski}, {Mohrmann}, {Montanari}, {Moore}, {Morris}, {Moulin}, {Muller},
  {Murach}, {Nakashima}, {Nayerhoda}, {de Naurois}, {Ndiyavala}, {Niemiec},
  {Oakes}, {O'Brien}, {Odaka}, {Ohm}, {Olivera-Nieto}, {de Ona Wilhelmi},
  {Ostrowski}, {Panny}, {Panter}, {Parsons}, {Peron}, {Peyaud}, {Piel}, {Pita},
  {Poireau}, {Priyana Noel}, {Prokhorov}, {Prokoph}, {P{\"u}hlhofer}, {Punch},
  {Quirrenbach}, {Raab}, {Rauth}, {Reichherzer}, {Reimer}, {Reimer}, {Remy},
  {Renaud}, {Rieger}, {Rinchiuso}, {Romoli}, {Rowell}, {Rudak}, {Ruiz-Velasco},
  {Sahakian}, {Sailer}, {Salzmann}, {Sanchez}, {Santangelo}, {Sasaki},
  {Scalici}, {Sch{\"a}fer}, {Sch{\"u}ssler}, {Schutte}, {Schwanke},
  {Seglar-Arroyo}, {Senniappan}, {Seyffert}, {Shafi}, {Shapopi},
  {Shiningayamwe}, {Simoni}, {Sinha}, {Sol}, {Specovius}, {Spencer},
  {Spir-Jacob}, {Stawarz}, {Sun}, {Steenkamp}, {Stegmann}, {Steinmassl},
  {Steppa}, {Takahashi}, {Tavernier}, {Taylor}, {Terrier}, {Thiersen},
  {Tiziani}, {Tluczykont}, {Tomankova}, {Trichard}, {Tsirou}, {Tuffs},
  {Uchiyama}, {van der Walt}, {van Eldik}, {van Rensburg}, {van Soelen},
  {Vasileiadis}, {Veh}, {Venter}, {Vincent}, {Vink}, {V{\"o}lk}, {Wadiasingh},
  {Wagner}, {Watson}, {Werner}, {White}, {Wierzcholska}, {Wun Wong},
  {Yusafzai}, {Zacharias}, {Zanin}, {Zargaryan}, {Zdziarski}, {Zech}, {Zhu},
  {Zorn}, {Zouari}, \& {{\.Z}ywucka}}]{hess_j1702}
{Abdalla}, H., {Aharonian}, F., {Ait Benkhali}, F., {et~al.} 2021, \aap, 653,
  A152

\bibitem[{{Aharonian} {et~al.}(2008){Aharonian}, {Akhperjanian}, {Barres de
  Almeida}, {Bazer-Bachi}, {Behera}, {Beilicke}, {Benbow}, {Bernl{\"o}hr},
  {Boisson}, {Bolz}, {Borrel}, {Braun}, {Brion}, {Brown}, {B{\"u}hler},
  {Bulik}, {B{\"u}sching}, {Boutelier}, {Carrigan}, {Chadwick}, {Chounet},
  {Clapson}, {Coignet}, {Cornils}, {Costamante}, {Dalton}, {Degrange},
  {Dickinson}, {Djannati-Ata{\"\i}}, {Domainko}, {Drury}, {Dubois}, {Dubus},
  {Dyks}, {Egberts}, {Emmanoulopoulos}, {Espigat}, {Farnier}, {Feinstein},
  {Fiasson}, {F{\"o}rster}, {Fontaine}, {Funk}, {F{\"u}{\ss}ling}, {Gallant},
  {Giebels}, {Glicenstein}, {Gl{\"u}ck}, {Goret}, {Hadjichristidis}, {Hauser},
  {Hauser}, {Heinzelmann}, {Henri}, {Hermann}, {Hinton}, {Hoffmann}, {Hofmann},
  {Holleran}, {Hoppe}, {Horns}, {Jacholkowska}, {de Jager}, {Jung},
  {Katarzy{\'n}ski}, {Kendziorra}, {Kerschhaggl}, {Kh{\'e}lifi}, {Keogh},
  {Komin}, {Kosack}, {Lamanna}, {Latham}, {Lemi{\`e}re}, {Lemoine-Goumard},
  {Lenain}, {Lohse}, {Martin}, {Martineau-Huynh}, {Marcowith}, {Masterson},
  {Maurin}, {Maurin}, {McComb}, {Moderski}, {Moulin}, {de Naurois}, {Nedbal},
  {Nolan}, {Ohm}, {Olive}, {de O{\~n}a Wilhelmi}, {Orford}, {Osborne},
  {Ostrowski}, {Panter}, {Pedaletti}, {Pelletier}, {Petrucci}, {Pita},
  {P{\"u}hlhofer}, {Punch}, {Ranchon}, {Raubenheimer}, {Raue}, {Rayner},
  {Renaud}, {Ripken}, {Rob}, {Rolland}, {Rosier-Lees}, {Rowell}, {Rudak},
  {Ruppel}, {Sahakian}, {Santangelo}, {Schlickeiser}, {Sch{\"o}ck},
  {Schr{\"o}der}, {Schwanke}, {Schwarzburg}, {Schwemmer}, {Shalchi}, {Sol},
  {Spangler}, {Stawarz}, {Steenkamp}, {Stegmann}, {Superina}, {Tam},
  {Tavernet}, {Terrier}, {van Eldik}, {Vasileiadis}, {Venter}, {Vialle},
  {Vincent}, {Vivier}, {V{\"o}lk}, {Volpe}, {Wagner}, {Ward}, {Zdziarski}, \&
  {Zech}}]{aharonian08}
{Aharonian}, F., {Akhperjanian}, A.~G., {Barres de Almeida}, U., {et~al.} 2008,
  \aap, 477, 353

\bibitem[{{Aharonian} {et~al.}(2006{\natexlab{a}}){Aharonian}, {Akhperjanian},
  {Bazer-Bachi}, {Beilicke}, {Benbow}, {Berge}, {Bernl{\"o}hr}, {Boisson},
  {Bolz}, {Borrel}, {Braun}, {Breitling}, {Brown}, {B{\"u}hler},
  {B{\"u}sching}, {Carrigan}, {Chadwick}, {Chounet}, {Cornils}, {Costamante},
  {Degrange}, {Dickinson}, {Djannati-Ata{\"\i}}, {O'C. Drury}, {Dubus},
  {Egberts}, {Emmanoulopoulos}, {Espigat}, {Feinstein}, {Ferrero}, {Fiasson},
  {Fontaine}, {Funk}, {Funk}, {Gallant}, {Giebels}, {Glicenstein}, {Goret},
  {Hadjichristidis}, {Hauser}, {Hauser}, {Heinzelmann}, {Henri}, {Hermann},
  {Hinton}, {Hofmann}, {Holleran}, {Horns}, {Jacholkowska}, {de Jager},
  {Kh{\'e}lifi}, {Komin}, {Konopelko}, {Kosack}, {Latham}, {Le Gallou},
  {Lemi{\`e}re}, {Lemoine-Goumard}, {Lohse}, {Martin}, {Martineau-Huynh},
  {Marcowith}, {Masterson}, {McComb}, {de Naurois}, {Nedbal}, {Nolan},
  {Noutsos}, {Orford}, {Osborne}, {Ouchrif}, {Panter}, {Pelletier}, {Pita},
  {P{\"u}hlhofer}, {Punch}, {Raubenheimer}, {Raue}, {Rayner}, {Reimer},
  {Reimer}, {Ripken}, {Rob}, {Rolland}, {Rowell}, {Sahakian}, {Saug{\'e}},
  {Schlenker}, {Schlickeiser}, {Schwanke}, {Sol}, {Spangler}, {Spanier},
  {Steenkamp}, {Stegmann}, {Superina}, {Tavernet}, {Terrier}, {Th{\'e}oret},
  {Tluczykont}, {van Eldik}, {Vasileiadis}, {Venter}, {Vincent}, {V{\"o}lk},
  {Wagner}, \& {Ward}}]{hess_crab}
{Aharonian}, F., {Akhperjanian}, A.~G., {Bazer-Bachi}, A.~R., {et~al.}
  2006{\natexlab{a}}, \aap, 457, 899

\bibitem[{{Aharonian} {et~al.}(2006{\natexlab{b}}){Aharonian}, {Akhperjanian},
  {Bazer-Bachi}, {Beilicke}, {Benbow}, {Berge}, {Bernl{\"o}hr}, {Boisson},
  {Bolz}, {Borrel}, {Braun}, {Breitling}, {Brown}, {Chadwick}, {Chounet},
  {Cornils}, {Costamante}, {Degrange}, {Dickinson}, {Djannati-Ata{\"\i}},
  {Drury}, {Dubus}, {Emmanoulopoulos}, {Espigat}, {Feinstein}, {Fontaine},
  {Fuchs}, {Funk}, {Gallant}, {Giebels}, {Gillessen}, {Glicenstein}, {Goret},
  {Hadjichristidis}, {Hauser}, {Heinzelmann}, {Henri}, {Hermann}, {Hinton},
  {Hofmann}, {Holleran}, {Horns}, {Jacholkowska}, {de Jager}, {Kh{\'e}lifi},
  {Komin}, {Konopelko}, {Latham}, {Le Gallou}, {Lemi{\`e}re},
  {Lemoine-Goumard}, {Leroy}, {Lohse}, {Martin}, {Martineau-Huynh},
  {Marcowith}, {Masterson}, {McComb}, {de Naurois}, {Nolan}, {Noutsos},
  {Orford}, {Osborne}, {Ouchrif}, {Panter}, {Pelletier}, {Pita},
  {P{\"u}hlhofer}, {Punch}, {Raubenheimer}, {Raue}, {Raux}, {Rayner}, {Reimer},
  {Reimer}, {Ripken}, {Rob}, {Rolland}, {Rowell}, {Sahakian}, {Saug{\'e}},
  {Schlenker}, {Schlickeiser}, {Schuster}, {Schwanke}, {Siewert}, {Sol},
  {Spangler}, {Steenkamp}, {Stegmann}, {Tavernet}, {Terrier}, {Th{\'e}oret},
  {Tluczykont}, {Vasileiadis}, {Venter}, {Vincent}, {V{\"o}lk}, \&
  {Wagner}}]{hgps1}
{Aharonian}, F., {Akhperjanian}, A.~G., {Bazer-Bachi}, A.~R., {et~al.}
  2006{\natexlab{b}}, \apj, 636, 777

\bibitem[{{Aharonian} {et~al.}(2007){Aharonian}, {Akhperjanian}, {Bazer-Bachi},
  {Beilicke}, {Benbow}, {Berge}, {Bernl{\"o}hr}, {Boisson}, {Bolz}, {Borrel},
  {Braun}, {Brion}, {Brown}, {B{\"u}hler}, {B{\"u}sching}, {Carrigan},
  {Chadwick}, {Chounet}, {Coignet}, {Cornils}, {Costamante}, {Degrange},
  {Dickinson}, {Djannati-Ata{\"\i}}, {O'C. Drury}, {Dubus}, {Egberts},
  {Emmanoulopoulos}, {Espigat}, {Feinstein}, {Ferrero}, {Fiasson}, {Fontaine},
  {Funk}, {Funk}, {F{\"u}{\ss}ling}, {Gallant}, {Giebels}, {Glicenstein},
  {Gl{\"u}ck}, {Goret}, {Hadjichristidis}, {Hauser}, {Hauser}, {Heinzelmann},
  {Henri}, {Hermann}, {Hinton}, {Hoffmann}, {Hofmann}, {Holleran}, {Hoppe},
  {Horns}, {Jacholkowska}, {de Jager}, {Kendziorra}, {Kerschhaggl},
  {Kh{\'e}lifi}, {Komin}, {Konopelko}, {Kosack}, {Lamanna}, {Latham}, {Le
  Gallou}, {Lemi{\`e}re}, {Lemoine-Goumard}, {Lohse}, {Martin},
  {Martineau-Huynh}, {Marcowith}, {Masterson}, {Maurin}, {McComb}, {Moulin},
  {de Naurois}, {Nedbal}, {Nolan}, {Noutsos}, {Olive}, {Orford}, {Osborne},
  {Panter}, {Pelletier}, {Pita}, {P{\"u}hlhofer}, {Punch}, {Ranchon},
  {Raubenheimer}, {Raue}, {Rayner}, {Reimer}, {Reimer}, {Ripken}, {Rob},
  {Rolland}, {Rosier-Lees}, {Rowell}, {Sahakian}, {Santangelo}, {Saug{\'e}},
  {Schlenker}, {Schlickeiser}, {Schr{\"o}der}, {Schwanke}, {Schwarzburg},
  {Schwemmer}, {Shalchi}, {Sol}, {Spangler}, {Spanier}, {Steenkamp},
  {Stegmann}, {Superina}, {Tam}, {Tavernet}, {Terrier}, {Tluczykont}, {van
  Eldik}, {Vasileiadis}, {Venter}, {Vialle}, {Vincent}, {V{\"o}lk}, {Wagner},
  \& {Ward}}]{hessj1713}
{Aharonian}, F., {Akhperjanian}, A.~G., {Bazer-Bachi}, A.~R., {et~al.} 2007,
  \aap, 464, 235

\bibitem[{{Aharonian} {et~al.}(2023){Aharonian}, {Malyshev}, \&
  {Chernyakova}}]{we_j1702}
{Aharonian}, F., {Malyshev}, D., \& {Chernyakova}, M. 2023, \apj, 955, 147

\bibitem[{{Aharonian} {et~al.}(2010){Aharonian}, {Kelner}, \&
  {Prosekin}}]{aharonian10}
{Aharonian}, F.~A., {Kelner}, S.~R., \& {Prosekin}, A.~Y. 2010, \prd, 82,
  043002

\bibitem[{{Chernyakova} {et~al.}(2011){Chernyakova}, {Malyshev}, {Aharonian},
  {Crocker}, \& {Jones}}]{Chernyakova2011}
{Chernyakova}, M., {Malyshev}, D., {Aharonian}, F.~A., {Crocker}, R.~M., \&
  {Jones}, D.~I. 2011, \apj, 726, 60

\bibitem[{{De Luca} \& {Molendi}(2004)}]{luca04}
{De Luca}, A. \& {Molendi}, S. 2004, \aap, 419, 837

\bibitem[{{Giunti} {et~al.}(2022){Giunti}, {Acero}, {Kh{\'e}lifi}, {Kosack},
  {Lemi{\`e}re}, \& {Terrier}}]{xmm22}
{Giunti}, L., {Acero}, F., {Kh{\'e}lifi}, B., {et~al.} 2022, \aap, 667, A130

\bibitem[{{Jansson} \& {Farrar}(2012)}]{jansson_farrar}
{Jansson}, R. \& {Farrar}, G.~R. 2012, \apj, 757, 14

\bibitem[{{Khangulyan} {et~al.}(2014){Khangulyan}, {Aharonian}, \&
  {Kelner}}]{khangulian14}
{Khangulyan}, D., {Aharonian}, F.~A., \& {Kelner}, S.~R. 2014, \apj, 783, 100

\bibitem[{{Malyshev} {et~al.}(2022){Malyshev}, {Moulin}, \&
  {Santangelo}}]{we_pbh}
{Malyshev}, D., {Moulin}, E., \& {Santangelo}, A. 2022, \prd, 106, 123020

\bibitem[{{Meidinger} {et~al.}(2020){Meidinger}, {Albrecht}, {Beitler},
  {Bonholzer}, {Emberger}, {Frank}, {Lederhuber}, {M{\"u}ller-Seidlitz},
  {Nandra}, {Oser}, {Ott}, {Plattner}, \& {Strecker}}]{athena_wfi}
{Meidinger}, N., {Albrecht}, S., {Beitler}, C., {et~al.} 2020, in Society of
  Photo-Optical Instrumentation Engineers (SPIE) Conference Series, Vol. 11444,
  Space Telescopes and Instrumentation 2020: Ultraviolet to Gamma Ray, ed.
  J.-W.~A. {den Herder}, S.~{Nikzad}, \& K.~{Nakazawa}, 114440T

\bibitem[{{Merloni} {et~al.}(2024){Merloni}, {Lamer}, {Liu}, {Ramos-Ceja},
  {Brunner}, {Bulbul}, {Dennerl}, {Doroshenko}, {Freyberg}, {Friedrich},
  {Gatuzz}, {Georgakakis}, {Haberl}, {Igo}, {Kreykenbohm}, {Liu}, {Maitra},
  {Malyali}, {Mayer}, {Nandra}, {Predehl}, {Robrade}, {Salvato}, {Sanders},
  {Stewart}, {Tub{\'\i}n-Arenas}, {Weber}, {Wilms}, {Arcodia}, {Artis},
  {Aschersleben}, {Avakyan}, {Aydar}, {Bahar}, {Balzer}, {Becker}, {Berger},
  {Boller}, {Bornemann}, {Br{\"u}ggen}, {Brusa}, {Buchner}, {Burwitz},
  {Camilloni}, {Clerc}, {Comparat}, {Coutinho}, {Czesla}, {Dannhauer},
  {Dauner}, {Dauser}, {Dietl}, {Dolag}, {Dwelly}, {Egg}, {Ehl}, {Freund},
  {Friedrich}, {Gaida}, {Garrel}, {Ghirardini}, {Gokus}, {Gr{\"u}nwald},
  {Grandis}, {Grotova}, {Gruen}, {Gueguen}, {H{\"a}mmerich}, {Hamaus},
  {Hasinger}, {Haubner}, {Homan}, {Ider Chitham}, {Joseph}, {Joyce},
  {K{\"o}nig}, {Kaltenbrunner}, {Khokhriakova}, {Kink}, {Kirsch}, {Kluge},
  {Knies}, {Krippendorf}, {Krumpe}, {Kurpas}, {Li}, {Liu}, {Locatelli},
  {Lorenz}, {M{\"u}ller}, {Magaudda}, {Mannes}, {McCall}, {Meidinger},
  {Michailidis}, {Migkas}, {Mu{\~n}oz-Giraldo}, {Musiimenta}, {Nguyen-Dang},
  {Ni}, {Olechowska}, {Ota}, {Pacaud}, {Pasini}, {Perinati}, {Pires},
  {Pommranz}, {Ponti}, {Poppenhaeger}, {P{\"u}hlhofer}, {Rau}, {Reh},
  {Reiprich}, {Roster}, {Saeedi}, {Santangelo}, {Sasaki}, {Schmitt},
  {Schneider}, {Schrabback}, {Schuster}, {Schwope}, {Seppi}, {Serim},
  {Shreeram}, {Sokolova-Lapa}, {Starck}, {Stelzer}, {Stierhof}, {Suleimanov},
  {Tenzer}, {Traulsen}, {Tr{\"u}mper}, {Tsuge}, {Urrutia}, {Veronica},
  {Waddell}, {Willer}, {Wolf}, {Yeung}, {Zainab}, {Zangrandi}, {Zhang},
  {Zhang}, \& {Zheng}}]{dr1}
{Merloni}, A., {Lamer}, G., {Liu}, T., {et~al.} 2024, \aap, 682, A34

\bibitem[{{Porter} {et~al.}(2008){Porter}, {Moskalenko}, {Strong}, {Orlando},
  \& {Bouchet}}]{isrf}
{Porter}, T.~A., {Moskalenko}, I.~V., {Strong}, A.~W., {Orlando}, E., \&
  {Bouchet}, L. 2008, \apj, 682, 400

\bibitem[{{Predehl} {et~al.}(2021){Predehl}, {Andritschke}, {Arefiev},
  {Babyshkin}, {Batanov}, {Becker}, {B{\"o}hringer}, {Bogomolov}, {Boller},
  {Borm}, {Bornemann}, {Br{\"a}uninger}, {Br{\"u}ggen}, {Brunner}, {Brusa},
  {Bulbul}, {Buntov}, {Burwitz}, {Burkert}, {Clerc}, {Churazov}, {Coutinho},
  {Dauser}, {Dennerl}, {Doroshenko}, {Eder}, {Emberger}, {Eraerds},
  {Finoguenov}, {Freyberg}, {Friedrich}, {Friedrich}, {F{\"u}rmetz},
  {Georgakakis}, {Gilfanov}, {Granato}, {Grossberger}, {Gueguen}, {Gureev},
  {Haberl}, {H{\"a}lker}, {Hartner}, {Hasinger}, {Huber}, {Ji}, {Kienlin},
  {Kink}, {Korotkov}, {Kreykenbohm}, {Lamer}, {Lomakin}, {Lapshov}, {Liu},
  {Maitra}, {Meidinger}, {Menz}, {Merloni}, {Mernik}, {Mican}, {Mohr},
  {M{\"u}ller}, {Nandra}, {Nazarov}, {Pacaud}, {Pavlinsky}, {Perinati},
  {Pfeffermann}, {Pietschner}, {Ramos-Ceja}, {Rau}, {Reiffers}, {Reiprich},
  {Robrade}, {Salvato}, {Sanders}, {Santangelo}, {Sasaki}, {Scheuerle},
  {Schmid}, {Schmitt}, {Schwope}, {Shirshakov}, {Steinmetz}, {Stewart},
  {Str{\"u}der}, {Sunyaev}, {Tenzer}, {Tiedemann}, {Tr{\"u}mper}, {Voron},
  {Weber}, {Wilms}, \& {Yaroshenko}}]{erosita}
{Predehl}, P., {Andritschke}, R., {Arefiev}, V., {et~al.} 2021, \aap, 647, A1

\bibitem[{{Thorpe-Morgan} {et~al.}(2020){Thorpe-Morgan}, {Malyshev},
  {Santangelo}, {Jochum}, {J{\"a}ger}, {Sasaki}, \& {Saeedi}}]{we_ths}
{Thorpe-Morgan}, C., {Malyshev}, D., {Santangelo}, A., {et~al.} 2020, \prd,
  102, 123003

\bibitem[{{Wilms} {et~al.}(2000){Wilms}, {Allen}, \& {McCray}}]{wilms_abund}
{Wilms}, J., {Allen}, A., \& {McCray}, R. 2000, \apj, 542, 914

\bibitem[{{Zabalza}(2015)}]{naima}
{Zabalza}, V. 2015, Proc.~of International Cosmic Ray Conference 2015, 922

\end{thebibliography}
\end{document}